\documentclass[longauth]{aa} 

\usepackage{graphicx}
\usepackage{txfonts}
\usepackage{url,hyperref}

\defcitealias{Fermi_1FGL1018}{\textit{Fermi}-LAT Collaboration 2012}
\defcitealias{Fermi_LS5039}{\textit{Fermi}-LAT Collaboration 2009}
\defcitealias{FermiLMC}{\textit{Fermi}-LAT Collaboration 2016}
\defcitealias{3FGL}{\textit{Fermi}-LAT Collaboration 2015}

\defcitealias{HESS_1FGL1018}{HESS Collaboration 2015b}
\defcitealias{HESS_LS5039}{HESS Collaboration 2006}
\defcitealias{LMCsurvey}{HESS Collaboration 2015a}
\defcitealias{N157B}{HESS Collaboration 2012}

\newcommand{\dgr}{\ensuremath{^\circ}}
\newcommand\ergs{\ensuremath{\mathrm{erg\,s}^{-1}}}
\newcommand\cms{cm$^{-2}$\,s$^{-1}$\xspace}%
\newcommand\cmstev{cm$^{-2}$\,s$^{-1}$\,TeV$^{-1}$\xspace}%

\newcommand{\Msun}{\ensuremath{M_{\odot}}}

\newcommand{\Lsun}{\ensuremath{L_{\odot}}}

\newcommand{\dem}{DEM~L241}
\newcommand{\cxou}{CXOU~053600.0$-$673507}

\newcommand{\CAL}{CAL~60}
\newcommand{\hessj}{HESS~J0536$-$675}
\newcommand{\pthree}{LMC~P3}

\newcommand{\ls}{LS~5039}
\newcommand{\jten}{1FGL~J1018.6$-$5856}
\newcommand{\psrb}{PSR~B1259$-$63/LS~2883}

\newcommand{\fermi}{{\it Fermi}-LAT}
\newcommand{\hess}{H.E.S.S.}
\newcommand{\xmm}{{\it XMM-Newton}}
\newcommand{\chandra}{{\it Chandra}}

\newcommand{\amean}{0.32\,AU}

\newcommand{\Lacc}{\ensuremath{L_\mathrm{acc}}}

\newcommand{\Mco}{\ensuremath{M_\mathrm{co}}}
\newcommand{\Rco}{\ensuremath{R_\mathrm{co}}}

\newcommand{\MsunYr}{\ensuremath{{\Msun\mathrm{/yr}}}}

\newcommand{\Lint}{$(1.4 \pm 0.2) \times 10^{35}$\,\ergs}
\newcommand{\Lon}{$(5 \pm 1) \times 10^{35}$\,\ergs}

\newcommand{\SigInt}{$6.4\,\sigma$}
\newcommand{\SigOn}{$7.1\,\sigma$}
\newcommand{\SigOff}{$3.3\,\sigma$}

\def\gr{$\gamma$-ray}
\def\grs{$\gamma$ rays}

\begin{document} 

   \title{Detection of variable VHE \gr\ emission from the extra-galactic \gr\ binary \pthree}
\titlerunning{Detection of variable VHE \gr\ emission from the extra-galactic \gr\ binary \pthree}



\authorrunning{H.E.S.S. Collaboration}

\author{\tiny H.E.S.S. Collaboration
\and H.~Abdalla \inst{1}
\and A.~Abramowski \inst{2}
\and F.~Aharonian \inst{3,4,5}
\and F.~Ait~Benkhali \inst{3}
\and E.O.~Ang\"uner \inst{21}
\and M.~Arakawa \inst{43}
\and M.~Arrieta \inst{15}
\and P.~Aubert \inst{24}
\and M.~Backes \inst{8}
% ...
}

\institute{
Centre for Space Research, North-West University, Potchefstroom 2520, South Africa \and 
Universit\"at Hamburg, Institut f\"ur Experimentalphysik, Luruper Chaussee 149, D 22761 Hamburg, Germany \and 
Max-Planck-Institut f\"ur Kernphysik, P.O. Box 103980, D 69029 Heidelberg, Germany \and 
Dublin Institute for Advanced Studies, 31 Fitzwilliam Place, Dublin 2, Ireland \and 
National Academy of Sciences of the Republic of Armenia,  Marshall Baghramian Avenue, 24, 0019 Yerevan, Republic of Armenia  \and
Yerevan Physics Institute, 2 Alikhanian Brothers St., 375036 Yerevan, Armenia \and
Institut f\"ur Physik, Humboldt-Universit\"at zu Berlin, Newtonstr. 15, D 12489 Berlin, Germany \and
University of Namibia, Department of Physics, Private Bag 13301, Windhoek, Namibia \and
GRAPPA, Anton Pannekoek Institute for Astronomy, University of Amsterdam,  Science Park 904, 1098 XH Amsterdam, The Netherlands \and
Department of Physics and Electrical Engineering, Linnaeus University,  351 95 V\"axj\"o, Sweden \and
Institut f\"ur Theoretische Physik, Lehrstuhl IV: Weltraum und Astrophysik, Ruhr-Universit\"at Bochum, D 44780 Bochum, Germany \and
GRAPPA, Anton Pannekoek Institute for Astronomy and Institute of High-Energy Physics, University of Amsterdam,  Science Park 904, 1098 XH Amsterdam, The Netherlands \and
Institut f\"ur Astro- und Teilchenphysik, Leopold-Franzens-Universit\"at Innsbruck, A-6020 Innsbruck, Austria \and
School of Physical Sciences, University of Adelaide, Adelaide 5005, Australia \and
LUTH, Observatoire de Paris, PSL Research University, CNRS, Universit\'e Paris Diderot, 5 Place Jules Janssen, 92190 Meudon, France \and
Sorbonne Universit\'es, UPMC Universit\'e Paris 06, Universit\'e Paris Diderot, Sorbonne Paris Cit\'e, CNRS, Laboratoire de Physique Nucl\'eaire et de Hautes Energies (LPNHE), 4 place Jussieu, F-75252, Paris Cedex 5, France \and
Laboratoire Univers et Particules de Montpellier, Universit\'e Montpellier, CNRS/IN2P3,  CC 72, Place Eug\`ene Bataillon, F-34095 Montpellier Cedex 5, France \and
IRFU, CEA, Universit\'e Paris-Saclay, F-91191 Gif-sur-Yvette, France \and
Astronomical Observatory, The University of Warsaw, Al. Ujazdowskie 4, 00-478 Warsaw, Poland \and
Aix Marseille Universit\'e, CNRS/IN2P3, CPPM, Marseille, France \and
Instytut Fizyki J\c{a}drowej PAN, ul. Radzikowskiego 152, 31-342 Krak{\'o}w, Poland \and
Funded by EU FP7 Marie Curie, grant agreement No. PIEF-GA-2012-332350,  \and
School of Physics, University of the Witwatersrand, 1 Jan Smuts Avenue, Braamfontein, Johannesburg, 2050 South Africa \and
Laboratoire d'Annecy-le-Vieux de Physique des Particules, Universit\'{e} Savoie Mont-Blanc, CNRS/IN2P3, F-74941 Annecy-le-Vieux, France \and
Landessternwarte, Universit\"at Heidelberg, K\"onigstuhl, D 69117 Heidelberg, Germany \and
Universit\'e Bordeaux, CNRS/IN2P3, Centre d'\'Etudes Nucl\'eaires de Bordeaux Gradignan, 33175 Gradignan, France \and
Oskar Klein Centre, Department of Physics, Stockholm University, Albanova University Center, SE-10691 Stockholm, Sweden \and
Wallenberg Academy Fellow,  \and
Institut f\"ur Astronomie und Astrophysik, Universit\"at T\"ubingen, Sand 1, D 72076 T\"ubingen, Germany \and
Laboratoire Leprince-Ringuet, Ecole Polytechnique, CNRS/IN2P3, F-91128 Palaiseau, France \and
APC, AstroParticule et Cosmologie, Universit\'{e} Paris Diderot, CNRS/IN2P3, CEA/Irfu, Observatoire de Paris, Sorbonne Paris Cit\'{e}, 10, rue Alice Domon et L\'{e}onie Duquet, 75205 Paris Cedex 13, France \and
Univ. Grenoble Alpes, CNRS, IPAG, F-38000 Grenoble, France \and
Department of Physics and Astronomy, The University of Leicester, University Road, Leicester, LE1 7RH, United Kingdom \and
Nicolaus Copernicus Astronomical Center, Polish Academy of Sciences, ul. Bartycka 18, 00-716 Warsaw, Poland \and
Institut f\"ur Physik und Astronomie, Universit\"at Potsdam,  Karl-Liebknecht-Strasse 24/25, D 14476 Potsdam, Germany \and
Friedrich-Alexander-Universit\"at Erlangen-N\"urnberg, Erlangen Centre for Astroparticle Physics, Erwin-Rommel-Str. 1, D 91058 Erlangen, Germany \and
DESY, D-15738 Zeuthen, Germany \and
Obserwatorium Astronomiczne, Uniwersytet Jagiello{\'n}ski, ul. Orla 171, 30-244 Krak{\'o}w, Poland \and
Centre for Astronomy, Faculty of Physics, Astronomy and Informatics, Nicolaus Copernicus University,  Grudziadzka 5, 87-100 Torun, Poland \and
Department of Physics, University of the Free State,  PO Box 339, Bloemfontein 9300, South Africa \and
Heisenberg Fellow (DFG), ITA Universit\"at Heidelberg, Germany  \and
GRAPPA, Institute of High-Energy Physics, University of Amsterdam,  Science Park 904, 1098 XH Amsterdam, The Netherlands \and
Department of Physics, Rikkyo University, 3-34-1 Nishi-Ikebukuro, Toshima-ku, Tokyo 171-8501, Japan
% ...
}

\offprints{H.E.S.S.~collaboration,
\protect\\\email{\href{mailto:contact.hess@hess-experiment.eu}{contact.hess@hess-experiment.eu}};
\protect\\\protect\footnotemark[1] Corresponding authors
}

   \date{Received 06/12/2017; accepted 04/01/2018}

 
  \abstract
   {
   Recently, the high-energy (HE, 0.1--100\,GeV) \gr\ emission from the object \pthree\ in the Large Magellanic Cloud (LMC) has been discovered to be modulated with a 10.3-day period, making it the first extra-galactic \gr\ binary.
   }
   {
   This work aims at the detection of very-high-energy (VHE, >100\,GeV) \gr\ emission and the search for modulation of the VHE signal with the orbital period of the binary system.
   }
   {
   \pthree\ has been observed with the High Energy Stereoscopic System (\hess); the acceptance-corrected exposure time is 100\,h.
   The data set has been folded with the known orbital period of the system in order to test for variability of the emission. 
Energy spectra are obtained for the orbit-averaged data set, and for the orbital phase bin around the VHE maximum.
}
   {
   VHE \gr\ emission is detected with a statistical significance of \SigInt. The data clearly show variability which is phase-locked to the orbital period of the system. Periodicity cannot be deduced from the \hess\ data set alone. The orbit-averaged luminosity in the $1-10$\,TeV energy range is \Lint. A luminosity of \Lon\ is reached during 20\% of the orbit. HE and VHE \gr\ emissions are anti-correlated. \pthree\ is the most luminous \gr\ binary known so far.
   }
   {}

   \keywords{Gamma rays: stars; binaries; Stars: massive}

   \maketitle
%

\section{Introduction}

More than 60\% of all stellar systems containing high-mass stars (spectral type B2 or earlier) are binary or multiple systems \citep{2013ARA&A..51..269D}. When the more massive of the stars in these systems ends its life in a supernova explosion, a binary system is left behind where a compact object, either a neutron star or a black hole, is orbiting the remaining star.
If these objects radiate most of their power at energies of more than 1\,MeV then they are called \gr\ binaries.
The \gr\ emission arises either from the interaction of a pulsar wind (driven by the rotational energy loss of a rotating neutron star) with the stellar wind, or from accretion of the stellar wind onto a black hole or neutron star. The companion stars in these systems are either O- or Be-type stars. Only six \gr\ binaries have been identified so far.
The nature of the compact objects is generally unknown, with the exception of \psrb\ where the detection of pulsed emission \citep{1992ApJ...387L..37J} shows that the compact object is a neutron star. A review of \gr\ binaries and their properties is given by \citet{2013A&ARv..21...64D,Dubus}.

In order to identify previously undetected \gr\ binaries, \citet{Fermi} performed a  search for periodic emission from the sources in the \fermi\ 3FGL catalogue \citep{3FGL}. They found that the high-energy (HE, 0.1--100\,GeV) \gr\ signal of \pthree, an unidentified \gr\ source located in the Large Magellanic Cloud (LMC) \citepalias{FermiLMC}, is periodic with a period of $10.301 \pm 0.002$~days. In this publication, the phase~zero of this system is defined to correspond to the maximum of the HE \gr\ emission at MJD $57410.25 \pm 0.34$. 
The position of \pthree\ is consistent with the one of the soft X-ray source \CAL\ \citep{1981ApJ...248..925L}. \citet{1985AJ.....90...43C} identified a star of spectral type O5~III(f) as the likely counterpart of this X-ray source. Subsequent X-ray observations with \xmm\ \citep{Bamba} and \chandra\ \citep{Seward} confirmed a point-like X-ray source which is named \cxou. \citet{Seward} already concluded from the variabilities of the X-ray flux and the radial velocity of Balmer absorption lines that this object is likely a binary system. \pthree\ is located in the supernova remnant \dem, making it the third X-ray binary found in an observable supernova remnant after SS433/W\,50 \citep{SS433} and SXP\,1062 \citep{SXP1062}.

Little is known about the orbital parameters of the system. The most precise measurement of the orbital period comes from the HE \gr\ emission. \citet{Fermi} also analysed the radial velocity of the star and found an orbital period of 10.1~days and a superior conjunction of the companion star at MJD $57408.61 \pm 0.28$. The mass function prefers a neutron star as the compact object for a wide range of inclinations, but a black hole cannot be ruled out. The X-ray and radio emission of this object is modulated with the 10.3-day period, but is out of phase with the \gr\ emission \citep{Fermi}.

The detection of periodic \gr\ emission from a system with an O5-type companion star and the \gr\ to X-ray luminosity ratio allow a clear classification of this object as a high-mass \gr\ binary \citep{Fermi}. It is the first such object discovered outside the Milky Way. With an HE \gr\ luminosity in the energy range from 200\,MeV to 100\,GeV of $2.5\times10^{36}\,$\ergs\ \citepalias{FermiLMC} it is also the most luminous \gr\ binary known so far.

\section{\hess\ observations and results}

The LMC has been observed extensively with the High Energy Stereoscopic System (\hess) since 2004. These observations led to the discovery of three individual very-high-energy (VHE, >100\,GeV) \gr\--emitting sources \citepalias{N157B,LMCsurvey}. 
\hess\ is a system of five Imaging Air Cherenkov Telescopes, located in the Khomas Highland of Namibia at an altitude of 1800\,m.
It is sensitive to \grs\ of energies from tens of \,GeV up to several tens of TeV. The arrival direction of individual \grs\ can be reconstructed with an angular resolution of better than $0.1\dgr$, and their energy is estimated with a relative uncertainty of 15\%.
The data discussed here were taken between 2004 and the beginning of 2016 and add up to a total observation time of 277\,h, almost 70\,h more than what was used in the previous publication of LMC sources \citepalias{LMCsurvey}.
After correcting for dead time and camera offset angles, the effective (on-axis equivalent) exposure time for \pthree\ is 100\,h. 
About 5\% of these observations were taken with the participation of the large \hess-II telescope. The data recorded with this telescope are ignored in this analysis in order to obtain a homogeneous data set.
The data were analysed using \textit{Model analysis} with high-resolution cuts \citep{Mathieu}, where the camera images are compared with a semi-analytical model using a log-likelihood minimisation technique. The results were cross-checked with an independent multi-variate analysis chain based on image parametrisation \citep{TMVA,Disp}. The background was estimated from rings around the on regions to generate the \gr\ images \citep[\textit{ring background}, ][]{BGmodels} and from test regions with similar offsets from the camera centre for the spectral analysis \citep[\textit{reflected background}, ][]{BGmodels}. Due to the large zenith and offset angles as well as the event selection cuts the energy threshold for this data set is 714\,GeV.

   \begin{figure}
   \centering
   \includegraphics[width=0.88\hsize]{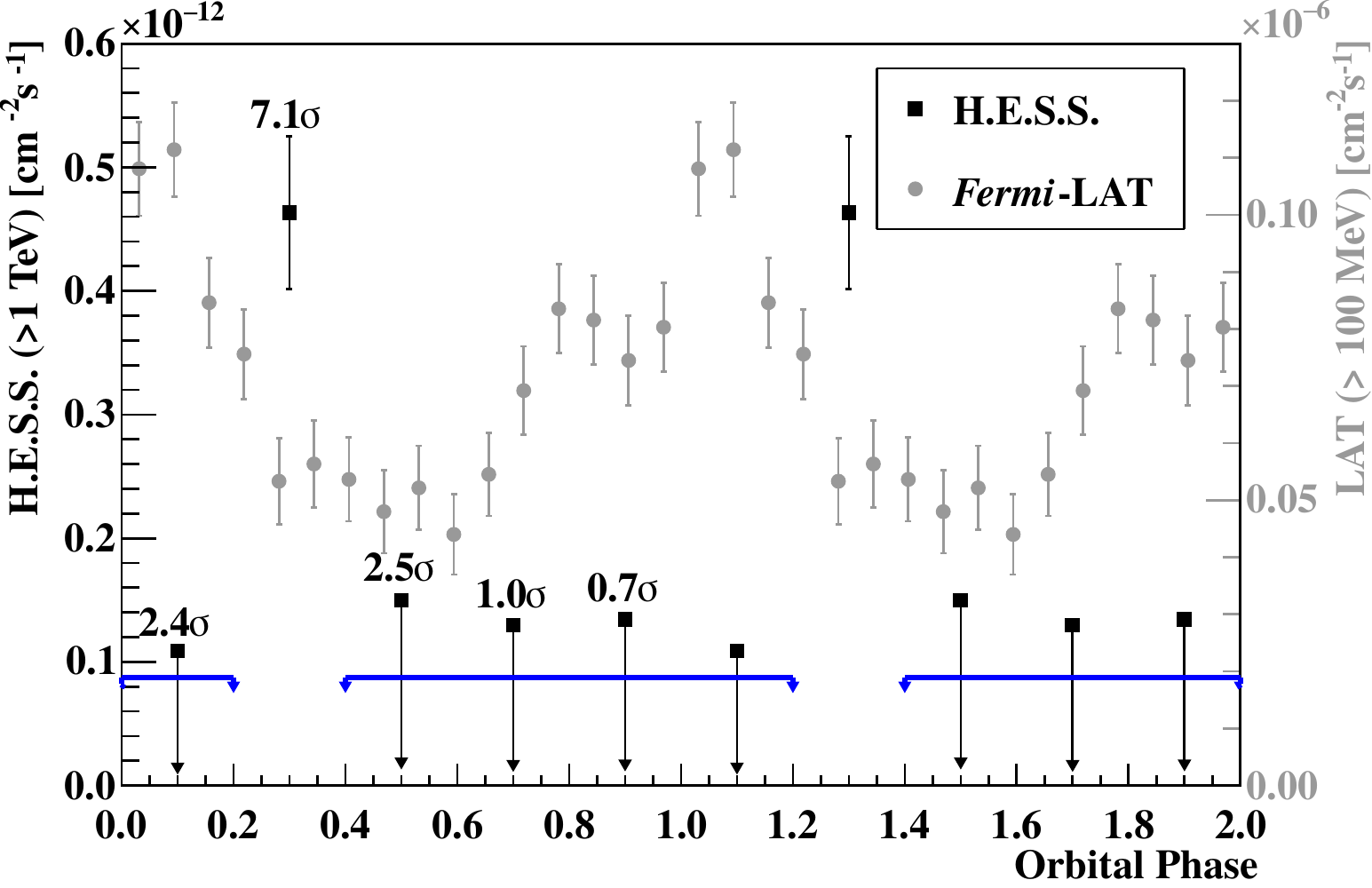}
      \caption{Folded \gr\ light curves with orbital phase zero at the maximum of the HE \gr\ emission (MJD~57410.25). For better readability two orbits are shown.
      The black points (left $y$-axis) represent the VHE light curve (this work), the grey points (right $y$-axis) represent the HE light curve \citep[from][, Fig.~3]{Fermi}.
      Error bars represent $1\,\sigma$ statistical uncertainty. For the phase bins without significant detection, upper limits (at 95\% confidence level) are given. The labels at the data points indicate the statistical significance of the excess in each phase bin. The blue horizontal lines denote the upper limit (at 95\% confidence level) on the flux in the off-peak region (orbital phase 0.4 to 1.2).}
         \label{fig:lightcurve}
   \end{figure}

   \begin{figure*}
   \centering
   \includegraphics[width=0.8\hsize]{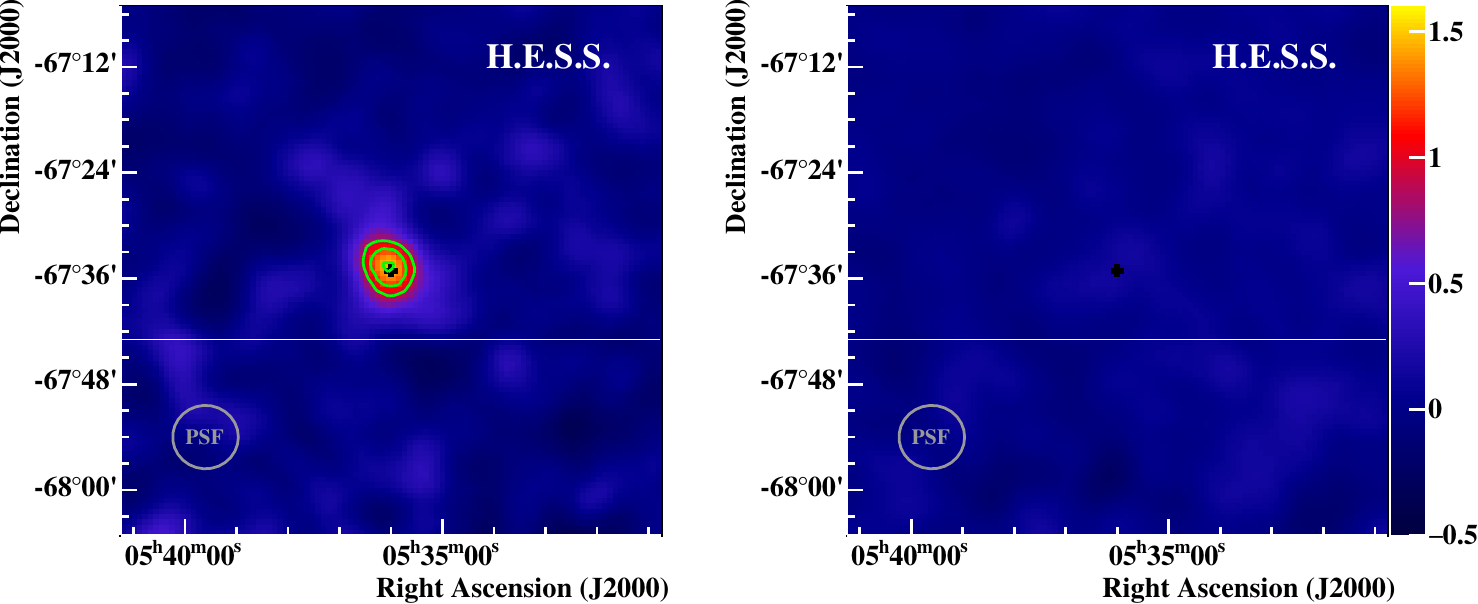}
   \caption{\hess\ excess count rate maps for the on-peak (left panel) and off-peak (right panel) regions of the orbit. The excess is smoothed with the point-spread function of the instrument (68\% containment radius of $0.06\degr$, indicated by the grey circle)
. The cross indicates the test position. Both plots have the same range, and the colour bar on the right-hand side is in units of excess counts per hour. The overlaid green contours represent 4, 5, and 6\,$\sigma$ statistical significance.
}
              \label{fig:skymap}%
    \end{figure*}

At the nominal position of \cxou\ an excess of 76.3 \gr\ events is detected with a statistical significance of \SigInt\ (Table~\ref{tab:results}).
The sensitivity of \hess\ does not allow a detection of flux variations of the object on a nightly basis. 
Therefore, the nightly light curve of the emission does not show any sign of variability; the fit of a constant yields $\chi^2 = 75.3$ for 99 degrees of freedom. The search for periodic emission using a Lomb-Scargle test \citep{Lomb,Scargle} and the Z-Transformed Discrete Correlation Function \citep{ZDCF} does not show significant periodicity.
Figure~\ref{fig:lightcurve} shows the light curve folded with the orbital period of the system of 10.301~days, where orbital phase zero is defined as the maximum of the HE light curve at MJD $57410.25$ \citep{Fermi}. Significant emission is detected only in the orbital phase bin between 0.2 and 0.4 with a pre-trial significance of \SigOn. 
This corresponds to a post-trial significance of $6.9\,\sigma$ after correcting for the test of five independent phase bins.
All other phase bins do not show significant emission (significances less than $2.5\,\sigma$, see Fig.~\ref{fig:lightcurve}). 
All phase bins have roughly the same exposure (between 18 and 21 hours).
Fitting the folded light curve with a constant results in a $\chi^2$ value of 27.03 for 4 degrees of freedom. The $\chi^2$ probability that the folded light curve is constant is hence less than $1.95 \times 10^{-5}$.
The emission is clearly variable and it is phase-locked to the orbital period of the system. Therefore, the detected VHE \gr\ emission can be associated to the binary system \pthree.
  
Figure~\ref{fig:skymap} shows the VHE \gr\ excess maps in the on-peak (orbital phase 0.2 to 0.4) and the off-peak (orbital phase 0.4 to 1.2) parts of the orbit. Fitting a point-like source folded with the instrument's point spread function results in a best-fit position of the source at RA = $5^{\mathrm{h}} 36^{\mathrm{m}} 0^{\mathrm{s}}$, Dec =
$-67^{\circ} 35\arcmin 11\arcsec$, equinox J2000, with a statistical uncertainty of $\pm 23\arcsec$ in each direction; the source is hence
labelled {\hessj}. The best-fit position is $4\arcsec$ (17\% of the statistical uncertainty) away from the nominal binary position. Therefore, the VHE \gr\ source is positionally compatible with the binary system.

   \begin{table*}
      \caption[]{Statistical results and spectral parameters for different orbital phase bins of \hessj. The on-peak region covers orbital phases from 0.2 to 0.4, the off-peak region the orbital phases from 0.4 to 1.2. Orbital phase zero is at the maximum of the HE \gr\ emission (MJD 57410.25).}
         \label{tab:results}
\begin{tabular}{lcccccc}
\hline
\hline
Orbital phase bin       & Excess & Significance 
                                        & $\Phi_{1\,\mathrm{TeV}}$ 
                                        & $ \Gamma$
                                        & $F (> 1\,\mathrm{TeV})$
                                        & $L (1 - 10\,\mathrm{TeV}, 50\,\mathrm{kpc})$ 
                                        \\
                                        &   &   
                                        & $[10^{-13}$ \cmstev$]$ 
                                        & 
                                        & $[10^{-13}$ \cms$]$
                                        & $[10^{35}$ \ergs$]$ 
                                        \\
\hline
full orbit      & 76.3
                        & \SigInt
                        & $2.0 \pm 0.4$
                        & $2.5 \pm 0.2$ 
                        & $1.4 \pm 0.4$
                        & $1.4 \pm 0.2$
                        \\
on-peak & 41.1
                        & \SigOn
                        & $5 \pm 1$
                        & $2.1 \pm 0.2$
                        & $5 \pm 2$
                        & $5 \pm 1$  
                        \\
off-peak        & 35.0
                        & \SigOff
                        & -
                        & $2.4$ (fixed) 
                        & $<0.88$ (95\% CL)
                        & $<0.88$ (95\% CL)
                        \\
\hline
\hline
\end{tabular}
   \end{table*}

%
   \begin{figure}
   \centering
   \includegraphics[width=0.88\hsize]{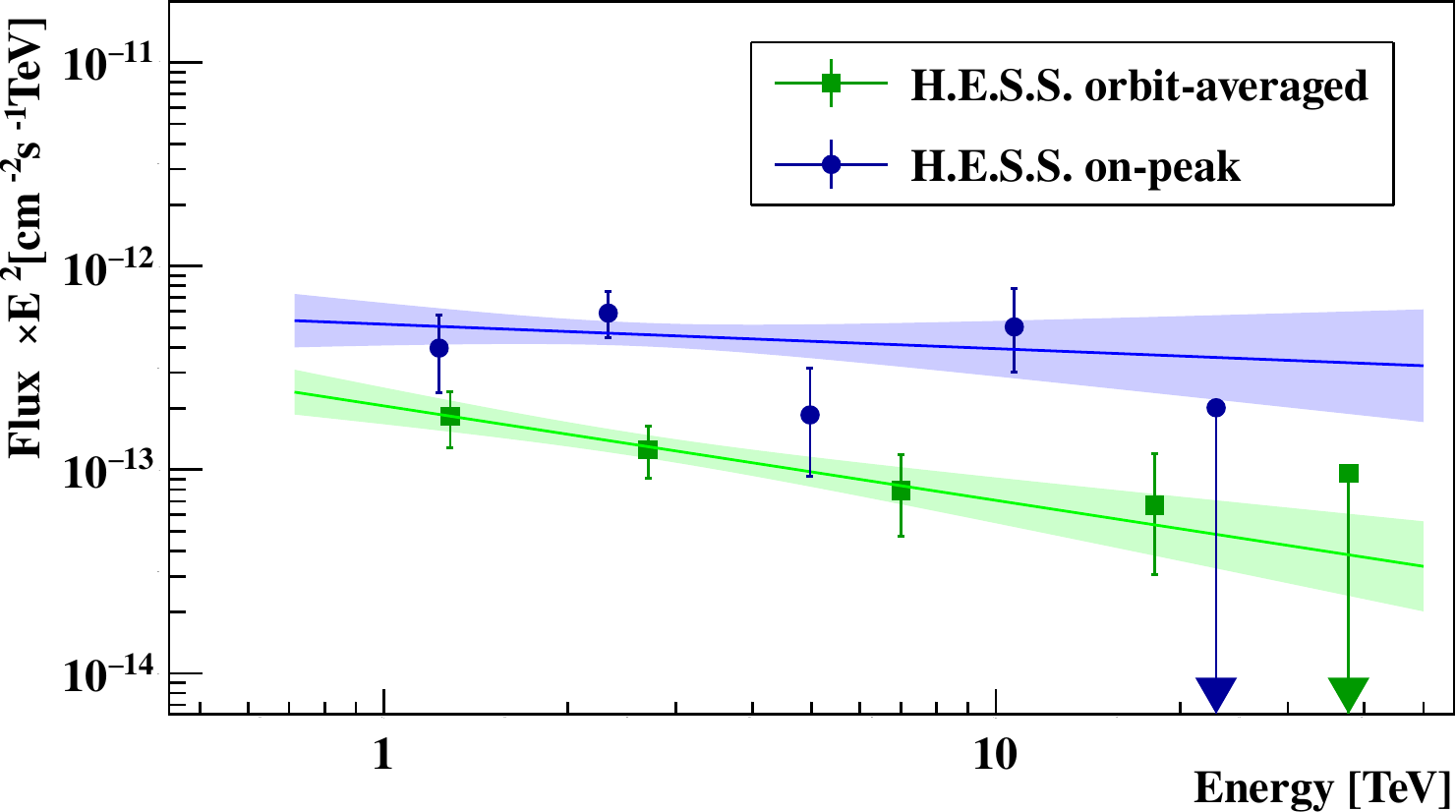}
      \caption{Spectral energy distribution averaged over the full orbit (green, squares) and for the on-peak orbital phase range (orbital phase from 0.2 to 0.4: blue, circles). The data points have $1\,\sigma$ statistical error bars, upper limits are for a 95\% confidence level. The best fit and its uncertainty are represented by the solid lines and shaded areas, respectively.
              }
         \label{fig:spectrum}
   \end{figure}


Figure~\ref{fig:spectrum} shows the spectral energy distribution for the full orbit and the on-peak part of the orbit. The spectra are fitted with a simple power law,
$\frac{dN}{dE} = \Phi_{1\,\mathrm{TeV}} \left( \frac{E}{1\,\mathrm{TeV}} \right)^{-\Gamma},$
and the best-fit parameters are summarised in Table~\ref{tab:results}. Significant emission is detected up to at least 10\,TeV and no high-energy cut-off of the spectra has been detected.
For the off-peak region of the orbit an upper limit on the integrated \gr\ flux has been obtained. Most of the \gr\ emission is radiated during only 20\% of the binary's orbit, when it reaches a flux of about four times the orbit-averaged flux. 

\section{Discussion}

The emission from \hessj\ is variable and modulated with the orbital period of the binary system, indicating that the size of the \gr\ emission region is at most the size of the binary system. The companion star in this system is of the type O5III, with a typical mass of 40\,\Msun\ \citep{Martins2005}. For a neutron star with a mass of 1.4\,\Msun\ as the compact object the semi-major axis of the orbit would be \amean.

The two scenarios proposed for \gr\ binaries are that the
\gr\ emission can be powered either by the spin-down of a pulsar or by accretion of the stellar wind onto the compact object. In the pulsar wind scenario, electrons are accelerated in the vicinity of the pulsar or in the shock front between the pulsar wind and stellar wind. The relativistic electrons produce \gr\ emission by inverse Compton (IC) upscattering of stellar photons. A typical O5III star has a surface temperature of 40\,kK and a luminosity of $5.4 \times 10^5\,\Lsun$ \citep{Martins2005}.
The IC scattering is in the Klein-Nishina regime and a hard electron spectrum with an index of $-1.5$ is required to produce the observed \gr\ spectrum.
For a binary separation of \amean\ the stellar photon field has an energy density of $253\,\mathrm{erg\,cm}^{-3}$. Electrons with energies between 0.5\, and 50 \,TeV and a total energy of $2.5\times10^{38}\,\mathrm{erg}$ are required to produce the observed \gr\ emission. The IC cooling time \citep[equation 1]{Khangulyan2008} of TeV electrons in this photon field is of the order of 100\,s. Therefore, the pulsar's spin-down power needs to be at least $10^{36}\,\ergs$ in order to provide the energy for the observed VHE \gr\ emission alone.

In the accretion scenario, a situation that is encountered in microquasars, the stellar wind of the companion star is accreted onto the compact object and gravitational potential energy is released as radiation.
The accretion luminosity of a neutron star is
\begin{equation}
\Lacc = 
\left( \frac{\dot{M} }{ 10^{-10} \MsunYr} \right) 
\left( \frac{\Mco}{ 1.4 \Msun } \right) 
\left( \frac{10 \mathrm{km} }{\Rco} \right) \times 
  1.2 \times 10^{36} \,\ergs,
\end{equation}
where \Mco\ and \Rco\ are the mass and radius of the compact object \citep{Accretion}. $\dot{M}$ is the mass-accretion rate which is of the order of $10^{-10} \MsunYr$ in a typical close binary system with accretion of the stellar wind. Accretion can power the observed \gr\ emission provided that the conversion efficiency from accretion power to \grs\ approaches unity.
Much higher mass-accretion rates and thus a higher accretion luminosity can be achieved by Roche lobe overflow, for instance an accretion rate of $10^{-4}\,\MsunYr$ is discussed for SS\,443 in such a scenario \citep[for a review see][]{Fabrika2004}. A more massive compact object can also increase the accretion luminosity by a factor of a few.

The peak of the VHE emission at phases between 0.2 and 0.4 coincides with the decline of the HE \gr\ emission towards its broad minimum between orbital phases 0.3 and 0.7 (see Fig.~\ref{fig:lightcurve}).
The variability of the \gr\ flux over the orbit of the system can be explained by the angle-dependent cross section of IC scattering, $\gamma\gamma$ absorption or the eccentricity of the orbit.
IC scattering is most efficient in head-on collisions in which the low-energy photons are backscattered. Therefore, the maximum of the \gr\ emission is expected at superior conjunction. Similarly, the minimum of  the \gr\ emission occurs at inferior conjunction, when only tail-on collisions are observed.
Further on, $\gamma\gamma$ absorption due to pair-production \citep{GouldSchreder1967} can modulate the VHE \gr\ emission with the orbital period and the flux maximum is expected at inferior conjunction \citep{BoettcherDermer2005}. 
This scenario is discussed for \ls\ \citepalias{Fermi_LS5039} and can naturally explain why the HE and VHE emissions of \pthree\ are  out of phase. In such a scenario the superior conjunction of \pthree\ would be around orbital phase 0.0 and inferior conjunction would be between 0.2 and 0.4.
The varying separation in an eccentric orbit could also explain the variability of the \gr\ emission. Near periastron, the compact object probes a higher stellar photon-field density and a higher stellar-wind density. In order to vary the \gr\ flux by a factor of around 4 (the ratio between the off-peak upper limit and the on-peak flux) by either IC emission or stellar-wind accretion, the binary separation must vary by a factor of 2, which is the case for an eccentricity of 0.33. In this case, and considering only VHE \gr\ emission, periastron would be between orbital phases 0.2 and 0.4.

HE and VHE \gr\ emission may also arise from two different particle populations. \citet{2013A&A...551A..17Z} propose the apex of the contact discontinuity between the compact object and the star as the source of the HE emission, and the pulsar wind termination shock on the opposite side as the source of the VHE emission. These emission sites together with geometric effects of the binary's orbit can also explain the phase-shift of the HE and VHE \gr\ light curves.

\pthree\ is the sixth, and the most luminous, \gr\ binary discovered so far.
With  an O-type companion star it is similar to \ls\ and \jten. 
With the HE and VHE emissions  out of phase, it resembles  \ls\ \citepalias{HESS_LS5039, Fermi_LS5039}, which is in contrast to \jten\ where HE and VHE emissions are in phase \citepalias{Fermi_1FGL1018, HESS_1FGL1018}.  But contrary to \ls, which shows VHE \gr\ emission during more than 50\% of the orbit, the duration of the emission of \hessj\ covers less than 20\% of the orbit (given the current instrument sensitivity).
The VHE light curve of \pthree\ is similar to the VHE \gr\ light curve of \jten\ \citepalias{HESS_1FGL1018}.
The detection of periodic HE and VHE emission from \pthree\ helps to fill the zoo of \gr\ binaries and will help to understand the underlying particle acceleration and \gr\ production mechanisms.
Detailed modelling of the HE and VHE light curves requires knowledge of the orbital parameters of the system.

The sensitivity of \hess\ only allows the detection of VHE \gr\ radiation during the high-state of the emission. The \SigOff\ statistical significance during the off-peak part of the orbit may indicate that the VHE emission extends into this part of the orbit as well. The ten-fold sensitivity of the future Cherenkov Telescope Array \citep[CTA,][]{2013APh....43....1H} will allow for the investigation of this part of the orbit of the binary system.

\section{Conclusions}


Variable VHE \gr\ emission from the newly discovered binary system \pthree\ has been detected with \hess\  The emission is phase-locked to the orbital period of the system. This makes \hessj\ the sixth VHE \gr\--emitting binary and the first extra-galactic \gr\ binary.
The energy spectrum of the VHE \gr\ emission is described by a simple power law. The orbit-averaged VHE luminosity of the system is \Lint. During 20\% of the orbit the VHE luminosity reaches \Lon. This makes \hessj\ the most luminous \gr\ binary known to date.
The VHE \gr\ emission can be either powered by the spin-down of a pulsar or by accretion of the stellar wind onto the compact object. A luminous pulsar ($\dot{E} \geq 10^{36}\,\ergs$), a high mass-accretion rate or a very massive compact object are needed to provide the energy for the observed VHE \gr\ emission.
The VHE emission is out of phase with the HE emission which may be explained by absorption due to pair production, or by different particle distributions responsible for the HE and VHE \gr\ production.
Observations with CTA may lead to detection of \gr\ emission during the off-peak part of the orbit allowing the modelling of the entire light curve.


\begin{acknowledgements}
The support of the Namibian authorities and of the University of Namibia in facilitating the construction and operation of \hess\ is gratefully acknowledged, as is the support by the German Ministry for Education and Research (BMBF), the Max Planck Society, the German Research Foundation (DFG), the Alexander von Humboldt Foundation, the Deutsche Forschungsgemeinschaft, the French Ministry for Research, the CNRS-IN2P3 and the Astroparticle Interdisciplinary Programme of the CNRS, the U.K. Science and Technology Facilities Council (STFC), the IPNP of the Charles University, the Czech Science Foundation, the Polish National Science Centre, the South African Department of Science and Technology and National Research Foundation, the University of Namibia, the National Commission on Research, Science \& Technology of Namibia (NCRST), the Innsbruck University, the Austrian Science Fund (FWF), and the Austrian Federal Ministry for Science, Research and Economy, the University of Adelaide and the Australian Research Council, the Japan Society for the Promotion of Science and by the University of Amsterdam.
We appreciate the excellent work of the technical support staff in Berlin, Durham, Hamburg, Heidelberg, Palaiseau, Paris, Saclay, and in Namibia in the construction and operation of the equipment. This work benefited from services provided by the H.E.S.S. Virtual Organisation, supported by the national resource providers of the EGI Federation.
We thank Robin Corbet for providing the HE light curve data shown in Fig.~\ref{fig:lightcurve}.
\end{acknowledgements}

   \bibliographystyle{aa} 
   \bibliography{CAL60_HESS} 

\end{document}